\documentstyle[editedvolume,namedreferences]{crckapbk}
\def \df {{\partial f}}

\def \ie {{\it ie \/}}

\def\pp{\parshape 2 0truecm 15truecm 2truecm 13truecm}
\def\apjref#1;#2;#3;#4;#5 {\par\pp#1, {\it #2}, {\bf #3}, #4. \par}
\def\ltsima{$\; \buildrel < \over \sim \;$}
\def\simlt{\lower.5ex\hbox{\ltsima}}
\def\gtsima{$\; \buildrel > \over \sim \;$}
\def\simgt{\lower.5ex\hbox{\gtsima}}
\def\kr{${\cal K}_r$}
\def\kz{${\cal K}_z$}

\def\kms{$\rm km s^{-1}$ }
\def\kms2{$\rm km s^{-2}$ }

\begin{opening}
\title{Astrometric Tests of Galactic Evolution
       }
\author{Gerard Gilmore}
\institute{Institute of Astronomy, Madingley Rd, Cambridge, UK\\
     gil@mail.ast.cam.ac.uk}

\end{opening}

\begin{document}

\section{Introduction}

There are many fundamental aspects of Galactic structure and evolution
which can be studied best or exclusively with high quality three
dimensional kinematics. Amongst these we note as examples determination
of the orientation of the stellar velocity ellipsoid, and the
detection of structure in velocity-position phase space. The first of
these is the primary limitation at present to reliable and accurate
measurement of the Galactic gravitational potential. The second is a
critical test of current standard models of Galactic formation and
evolution.

\section{Measuring Gravitational Potentials}

The classical method of measuring mass utilises the motions of stellar
tracers in the Galaxy. This essentially measures the
gravity-pressure-angular momentum balance of a suitable
dynamically-relaxed tracer population. Appropriate analysis allows the
gravitational force to be derived, and from this, {\it via} Poisson's
equation, the mass density that generates that potential. All such
analyses are applications of the collisionless Boltzmann equation. All
are in practice approximate.  The primary limitation in present
analyses is the absence of accurate 3-dimensional kinematics, of the
type which would be provided by a high precision astrometric project.

\subsection{The Collisionless Boltzmann Equation}
The dynamics of any large stellar system are governed by the collisionless
Bolzmann equation
\begin{equation}
\frac{D\! f}{Dt} \equiv \frac{\df}{\partial t} +
\frac{\partial\vec{x}}{\partial t}\cdot\frac{\df}{\partial \vec{x}} +
\frac{\partial\vec{v}}{\partial t}\cdot\frac{\df}{\partial \vec{v}} = 0,
\end{equation}
where $f$ is the phase space density at the point $(\vec{x},\vec{v})$ in phase
space (\ie\ there are $f(\vec{x},\vec{v})d^3\vec{x} d^3\vec{v} $ stars in a
volume of size $d^3\vec{x}$ centered on $\vec{x}$ with velocity in the
volume of size $d^3\vec{v}$ about $\vec{v}$).

The collisionless Boltzmann equation is satisfied by {\em any} stellar
population, whether other stars are present or not. This arises
because stars do not interact except through long-range gravity
forces, and those are being described through a smooth background
potential. Consequently, $f$ does not have to describe the entire
Galaxy; one can concentrate on any subsample of stars, and apply the
collisionless Boltzmann equation to it. Such subsamples are {\em
tracer populations}, since one may use their kinematics to trace the
potential of the Galaxy, irrespective of what generates this
potential.

If one has a steady-state tracer population, and a time-independent
potential, as the large-scale field in the Milky Way apparently is to
an adequate approximation for the present purpose, then
\begin{equation}
\frac{\df}{\partial t}=0.
\end{equation}
While the Galaxy is not rotationally symmetric, the essential features
of the present analysis do not depend on this asymmetry, so it may be
suppressed for now. It is
convenient to write out the collisionless Boltzmann equation in cylindrical
polar coordinates $(r,\phi,z)$ in which $z=0$ is the disk plane of
symmetry, with corresponding velocity components $(v_r,v_\phi,v_z)$:
\begin{equation}
v_r\frac{\df}{\partial r}
+ v_z\frac{\df}{\partial z}
+ \left({\cal K}_r+\frac{{v_\phi}^2}{r}\right) \frac{\df}{\partial v_r}
- \frac{v_r v_\phi}{r} \frac{\df}{\partial v_\phi}
+ {\cal K}_z \frac{\df}{\partial v_z}
=0
\end{equation}
where the accelerations $\dot v_r,\dot v_\phi,\dot v_z$ explicit in
the Boltzmann equation  have been equated to the forces (real and fictitious)
that cause them, and $\phi$-gradients in $f$ and in the potential set to
zero. The vector $\vec{\cal K}(r,z)$ is the gravity force.
Then clearly knowledge of $f(\vec{x},\vec{v})$ allows the force components \kr\
and \kz\ to be derived. Note, though, that a general function $f$ will not
allow a unique solution for \kr\ and \kz : $f$ has five independent variables
(we
suppressed any $\phi$-dependence) and so cannot in general be made to
satisfy the equation above, which contains only two
functions of two variables. Since the equation cannot easily
be solved in general with real data, one simplifies the analysis and
proceeds by taking velocity moments. Multiplying through by $v_z$ and by
$v_r$ and integrating over all velocity space produces Jeans' equations:
\begin{eqnarray}
\nu{\cal K}_z &=& \frac{\partial}{\partial z}(\nu\sigma^2_{zz}) +
               \frac{1}{r} \frac{\partial}{\partial r}(R\nu\sigma^2_{rz})\\
\label{eq-asdrift}
\nu{\cal K}_r &=& \frac{1}{r}\frac{\partial}{\partial r}(r\nu\sigma^2_{rr}) +
               \frac{\partial}{\partial z}(\nu\sigma^2_{rz}) -
               \frac{\nu\sigma^2_{\phi\phi}}{r},
\end{eqnarray}
where $\nu(r,z)$ is the space density of the stars, and
$\vec{\vec{\sigma}}(r,z)$ their velocity dispersion
tensor (\ie\ $\sigma^2_{ij}=\left\langle v_i v_j\right\rangle$).
In this way we have separated the two force components (radial and
vertical in this case), and can in principle derive them both from
measurements of the moments of the velocity distributions and density
of a stellar tracer population.

\subsection{The Tilt Term}

The term involving $\sigma_{rz}$ in the Jeans equation is related to
the tilt of the velocity ellipsoid in the (r,$z$)
plane. It is our lack of knowledge of this term which is the greatest
limitation in determination of mass distributions in the Galaxy. Two
specific examples will suffice to illustrate this.
This term has been treated explicitely in derivation of the surface
mass density and scale length of the Galactic disk (Kuijken \& Gilmore
1989abc,1991; Fux \& Martinet 1994; Cuddeford \& Amendt 1992), which
utilize eqn (4). It has also been discussed in terms of eqn (5), for
the disk asymmetric drift by Gilmore, Wyse \& Kuijken (1989), and when
relating the shape of the stellar distribution in the galactic halo to
the shape of the distribution of dark matter by (for example) van der
Marel (1991).

The two limiting cases may be discussed analytically, and obviously
include orientation of the stellar velocity ellipsoid in cylindrical
or in spherical polar coordinates. As an example, Kuijken \& Gilmore
assumed the long axis of the velocity ellipsoid
always points towards the centre of
the Galaxy, as is appropriate for a round potential.  In this case, they
could treat the $\sigma_{rz}$ term as an extra force term.  If the
Galaxy's disk, like those of other spirals, has a vertical scale
height which is constant with radius and a radially exponential
surface density profile $\mu \propto e^{-r/{\rm {h_r}}}$, (see Fux \&
Martinet 1994) then
$\sigma_{zz}\propto \mu \propto e^{-r/{\rm {h_r}}}$, and $\nu\propto
\mu \propto e^{-r/{\rm {h_r}}}$. from this Kuijken \& Gilmore were
able to derive a relationship between $\sigma_{rz}$ and $\sigma_{zz}$.
Since the details of this derivation have not previously been
published, we present them here.

\subsection{Derivation of the Tilt Term $\sigma_{rz}$}

Define the coordinate frames as:

\noindent Cylindrical polar $(r,z,\phi)$, with $\phi$ suppressed by
symmetry, and with velocity components $\dot z$ and $\dot r$;

\noindent Spherical polar, (R,$\theta, \phi$), where
we make $\theta$  the angle subtended at the Galactic centre between the
radius vector R and the radial planar coordinate $r$, so that
$tan\theta = z/r$, and suppress $\phi$ by symmetry. The relevant
velocity components, to avoid too many levels of subscripts, are $(A,B,C)$,
with $B$ suppressed by symmetry.
Thus, the $A$ component of velocity is the radial velocity away from the
Galactic centre, the $C$ component is perpendicular to it, and at $z=0$
one has $\dot r \equiv A$, and $\dot z \equiv C$.

The coordinate transformations by simple trigonometry are:
\begin{eqnarray} {\dot z} &=& { A sin\theta + C cos\theta,}\\
{\dot r} &=& { A cos\theta - C sin\theta}
\end{eqnarray}
By definition of the orientation of the velocity ellipsoid, Cov$(A,C)=0$,
and
\begin{eqnarray} \sigma_{rz} &=& {\rm Cov}(\dot r,\dot z)\\
 &=& (\dot z - \langle\dot z\rangle)(\dot r - \langle\dot r\rangle)\\
 &=& {\dot z}{\dot r}\\
{\dot z}{\dot r}
&=& (A sin\theta + C cos\theta)(A cos\theta - C sin\theta)\\
&=& A^2 sin\theta cos\theta - C^2 cos\theta sin\theta \,\, (+AC\,\,
terms)\\
\Longrightarrow \sigma_{rz}
&=& sin\theta cos\theta(\sigma_{AA} - \sigma_{CC})\\
{\dot z}{\dot z}
&=& (A sin\theta + C cos\theta)(A sin\theta + C cos\theta)\\
&=& A^2 sin^2\theta + C^2 cos^2\theta \,\, (+AC\,\, terms)\\
\Longrightarrow \sigma_{zz}
&=& \sigma_{AA} sin^2\theta + \sigma_{CC}cos^2\theta
\end{eqnarray}

Defining $\displaystyle \sigma_{AA} = \alpha^2\sigma_{CC}$ gives
\begin{equation}
\sigma_{zz} = (\alpha^2 sin^2\theta + cos^2\theta)\sigma_{CC}
\end{equation}
and
\begin{eqnarray}
\sigma_{rz} &=&
sin\theta cos\theta(\alpha^2 - 1)\sigma_{CC}\\
&=&{{sin\theta cos\theta (\alpha^2 - 1)} \over {\alpha^2sin^2\theta +
cos^2\theta}}\sigma_{zz}
\end{eqnarray}
(Note: At
$z=0$ of course $A \equiv \dot r$, and $C \equiv \dot z$, so
$\displaystyle \sigma_{rr} = \alpha^2\sigma_{zz}$, but this cannot be
true in general.)

Since tan$\theta = z/r, {\rm sin}\theta \propto z, {\rm cos}\theta
\propto r$, and hence
\begin{equation}
\sigma_{rz} = {{rz(\alpha^2 - 1)} \over {(\alpha^2 z^2 + r^2)}}
\sigma_{zz}
\end{equation}

This then leads directly to:
\begin{eqnarray}
{1\over \nu r}
{\partial\over\partial r} {(r\nu \sigma_{rz})} &=&6\sigma_{zz}
{\left\lbrace {4z^3\over{(4z^2+r^2)}^2} - {rz\over {\rm
h_r}(4z^2+r^2)} \right\rbrace}\\ { }&=&{\rm T}(r,z) \sigma_{zz}
\end{eqnarray}
The vertical Jeans' equation now becomes
\begin{equation}
{\cal K}_{z,\rm
eff}={\cal K}_z - {\rm T}(r,z) \sigma_{zz} ={1\over\nu}
{\partial\over\partial z} (\nu\sigma_{zz}).
\end{equation}
Given a `true' \kz, this linear equation can be solved for
$\sigma_{zz}$ (especially since T has an analytic $z$-integral), and
hence the effective force ${\cal K}_{z,{\rm eff}}$ calculated, with
its corresponding potential.  Kuijken \& Gilmore were then able to
proceed by assuming, as a first approximation, that the tracer
population moved under the action of this potential.

Given the quality
of observational data currently available relevant to the
determination of the Galactic \kz force law, the validity or otherwise
of this assumption is the primary uncertainty in present
determinations of the mass distribution in the optical parts of the
Galaxy. The true orientation of the stellar ellisoid could be measured
directly from suitable precise astrometric data. Data for a large
number of stars at distances of up to 2kpc, with a  space motion
precision of about 1km/s are required. Sub-milliarcsec astrometry can
realize this precision. We would then have a
precise, reliable, and assumption-free direct determination of that part
of the total Galactic mass density associated with the Galactic disk, and
that part distributed in a (dark) halo.

\section{The Shape of the Dark Halo}

Arguments similar to those above, but applied to the radial Jeans
equation (5) have been outlined by van der Marel (1991). He applied
the methodology to determination of the shape of the dark matter
distribution. The observational constraints are radial velocity data
at several locations and star count determinations of the shape of the
stellar distribution. The most recent and extensive determination of
the shape of the stellar halo, which is in good agreement with most
earlier determinations, provides $c/a=0.55$ (Larsen \& Humphries 1994).

Van der Marel shows that this observed shape of the stellar halo
implies a corresponding axis ratio $c/a\approx 0.25$ for the case of
an adopted spherical alignment of the velocity ellipsoid, and an axis
ratio $c/a \approx 0.55$ for the case of an adopted cylindrical
alignment of the stellar velocity ellipsoid. That is, there is more
than a factor of two uncertainty generated by the absence of 3-D
kinematics. This fractional error has the same effect as an error of a
factor of two in the observed shape of the stellar halo, an error
bound substantially outside current observational limits. That is, our
best determinations of the shape of the dark matter halo in the Galaxy
from a mix of stellar distribution and (radial velocity) kinematic
data are entirely dominated by missing suitably precise
astrometric data. These limits
could be removed by provision of data with a precision of a few km/s
for stars at distances of 5-10kpc.

\section{Tests of Galactic Formation and Evolution}

Modern models of Galaxy formation make fairly specific predictions
which are amenable to detailed test with Galactic kinematic and
chemical abundance data.  For example, popular Cold Dark Matter models
`predict' growth of the Galaxy about a central core, which should
contain the oldest stars. Later accretion of material forms the outer
halo and the disks, while continuing accretion will continue to affect
the kinematic structure of both the outer halo and the thin disk.

This blend-and-stir process will have been common at high redshifts,
when a rain of dwarf `proto-galaxies' was normal weather for a budding
giant. It continues today, at a rate which may still be significant
for some galaxies. The term `significant' here is worth some thought:
in the central regions of galaxies masses are large, timescales are
short and dynamical friction effective. Thus significant changes to a
galaxy require mergers of components of comparable mass. It has been
suggested that such mergers would destroy the thin disk of a galaxy
like the Milky Way, as argued recently by Toth \& Ostriker(1992). If
this argument were correct then normal late-type spirals must have
completed the bulk of their merger events at very early times.

In the outer parts of galaxies  mass densities are low, the fraction
of the total luminous galaxy which is seen is very small, and
timescales are comparable to a Hubble time. Thus one expects
relatively little fossil kinematic structure to be observble in the central
regions of normal galaxies, but it is probable that a large fraction
of the outer parts of a large galaxy is a recent (on kinematic
timescales) acquisition from afar. Fundamentally, the central regions
of a galaxy need not be related in any obvious way to the outer parts of
that same galaxy.

This picture, which contains aspects of both the monolithic (`ELS')
and the multi-fragment (`Searle-Zinn') pictures often discussed in
chemical evolution models, makes some specific predictions which are
amenable to test.  One specific example of current interest is the
`prediction' that mergers of small satellites are an essential feature
of galactic evolution.  This leads one to look for kinematic and
spatial structures, and `moving groups', as a primary test of such
models. That is, the galaxy formation concepts outlined above suggest
that a considerable amount of structure in the phase space
distribution function for those stars (and DM particles) which inhabit
the outer reaches of the galaxy is to be expected. This structure will
be the remnant dispersion orbits occupied by the debris of former
galactic satellites and near-neighbours which have now lost their
former isolated identity. The existence of this structure provides a
challenge in two ways: to devise dynamical analysis methods and/or
sample selection methods which will still provide a `fair sample' of
the outer galaxy for dynamical studies; and to identify the fractional
amount of phase space substructure, if any, and so test the (CDM)
galaxy merger models.

The existence of such phase space structure in stars of the thick disk
and halo, in addition to the younger stellar populations, has been
persuasively argued by Eggen for many years (cf Eggen 1987 for a review).  Such
moving
groups are of course a specific high contrast example of the structure
being considered here.

Direct evidence for an ongoing merger event has been discovered
recently in a study by Ibata, Gilmore \& Irwin (1994).  While
investigating the kinematic structure of the Galactic bulge, they
discovered a large phase-space structure consisting of $>100$ K giant,
M giant and carbon stars in three low Galactic latitude fields. The
group has a velocity dispersion of $<10$ kms$^{-1}$, and a mean
heliocentric radial velocity of 140 kms$^{-1}$, that varies by less
than 5 kms$^{-1}$ over the $8^\circ$ wide region of sky that the three
kinematic fields cover.  In a subsequent determination of the
colour-magnitude relationship for that line of sight, a giant branch,
red horizontal clump and horizontal branch are clearly visible,
superimposed on the general distribution made up of stars in the
bulge, and in the Galactic foreground. Stars
belonging to the low velocity dispersion group lie on the upper giant
branch of the colour-magnitude relation.  From the magnitude of the
horizontal branch, Ibata, Gilmore \& Irwin find that the object is
situated $15 \pm 2$ kpc from the Galactic centre.

An isodensity map shows an object which is elongated (with axial ratio
$\approx 3$), spanning $>10^\circ$ on the sky in a direction
perpendicular to the Galactic plane. The interpretation is clear: this
is a discovery, in kinematic phase space, of a dwarf (former) satellite galaxy
currently well inside the galactic optical boundary.
The tidal radius of the dwarf galaxy is then approximately an order of
magnitude smaller than its apparent size on the sky, so most of its
members will disperse into the Galactic halo over the next $\approx
10^8$ years. This finding clearly supports galaxy formation scenarios
of with significant merging events happening right up to the present
epoch.

The implications for the present are substantial. Kinematics can and
has discovered phase space structure; presumably much more remains to
be found. Kinematics can determine the present and future orbits of
the (former) member stars of this dwarf galaxy. Kinematics can map out
the merger history of the Milky Way. When astrometric data can provide
distances to a few percent, and kinematics to a few km/s, at distances
up to 20kpc from the Sun, then we will be able to determine in detail
the evolutionary history, and the three-dimensional distribution of
mass, in the Galaxy.

\end{document}